\documentclass[smallcondensed]{svjour3}

\smartqed
\usepackage{natbib}
\usepackage{mathptmx}
\usepackage{amsmath, amssymb, amsfonts}
\usepackage{graphicx}
\usepackage{epstopdf}
\usepackage{epsfig,color}
\usepackage[utf8]{inputenc}

\newcommand{\tr}{^{\prime}}
\def\bd#1{\mbox{\boldmath $#1$}}

\journalname{Quality\&Quantity}

\begin{document}
\title{Estimation of voter transitions and the Ecological fallacy}
\titlerunning{Ecological Bias}
\author{Antonio Forcina \and Davide Pellegrino}
\institute{A. Forcina \at
              Dipartimento di Economia \\
              via Pascoli, 06100 Perugia, Italy \\
              \email{forcinarosara@gmail.com}
              \and D. Pellegrino \at
              Dipartimento interateneo di
              Scienze, Progetto e Politiche del Territorio,\\
              Università e Politecnico di Torino, \\
              viale Mattioli 39, 10125 Torino, Italy
              }
\date{00.00.0000}
\maketitle
\begin{abstract}
This paper attempts an investigation into the features of ecological fallacy in the context of estimation of voter transitions between two elections. After reviewing some theoretical findings not always well understood, we discuss two tools for checking whether bias is present: (i) fitting models with covariates; (ii) comparing the standard errors of transition probabilities computed under ideal conditions against those based on bootstrap methods. Concerning the effect of covariates, we provide theoretical arguments and empirical evidence to show that, under certain conditions, modelling the effect of covariates may fail to correct ecological bias. Our investigation relies on the analysis of real and artificial data sets: the latter are obtained by a computer software which mimics voting behaviour. An application to a recent election in the city of Turin is also used to illustrate our methodology and findings.

\end{abstract}
\section{Introduction}
When trying to interpret the results of an election, substantial additional insights are available if estimates of voter transitions relative to a previous election are available. For example we may measure the proportion of faithful voters within different parties, the size of protest voters moving to minor parties or to abstention or the size and direction of strategic voting \citep{Herrmann} when the supporters of candidates not admitted to the runoffs, have to choose whether to abstain or support one of the competing candidates.

In order to compute voter transitions we need an estimate of the joint distribution of voters' decisions in the two elections under consideration. If available, these data could be summarised in an $R\times C$ contingency table with the voting options in the previous election by row and those of the new election by column. Unfortunately, due to the special nature of electoral data, the only information available is  the distribution of voters, separately, in each election.

Following the usual terminology, we call {\it aggregate data} the frequency distributions of voters by local unit and voting options, separately, in each election. We also call {\it individual data} the frequency distribution of voters according to the pair of voting options selected in the two elections under examination.
For the purpose of our analysis it is important to distinguish between the {\it full} and the {\it condensed} versions of individual data, according to whether the joint distributions are available within each local unit or only for the whole area under investigation.

Condensed individual data may be available in the context of simultaneous elections like in the case examined by \cite{Plescia2018}. Estimates could be obtained from exit polls, however, as discussed, for instance, in \cite{Russo14}, these estimates may be seriously biased due to non response, voters not telling the truth or not remembering their earlier decision.  There are three additional limitations with exit polls:
\begin{itemize}
\item it may be difficult to sample from abstainers as they do not show up at the polls;
\item while a sample size of a few thousands may provide accurate estimates of the proportion of voters for each party in a given election, a much larger sample size is necessary to estimate how the voters of a small party split among the options available in the new election;
\item often, voter transitions have strong local features which would cancel out when sampled on a whole region or State.
\end{itemize}
Based on a comparison with estimates provided from individual data, \cite{Liu07} noted that, with his data, estimates from exit polls were even worse that those provided by ecological regression.

Due to these limitations, in most situations ecological inference is the only feasible approach for the estimation of voter transitions. Thus it may be useful to study in depth, in the context of electoral data, under which conditions these estimates may be biased, a phenomenon known as ecological fallacy. A brief history of the subject starts with the seminal paper by \cite{Robinson}; substantial contributions were also provided by \cite{Firebaugh} and, more recently, \cite{Wakefield}. By elaborating on the last two papers, \cite{Gnaldi18} provide a more explicit characterization of the conditions for having  ecological fallacy; these results will be the starting point of the present paper which relies on the analysis of real and artificial electoral data generated according to a well defined model of voting behaviour presented in section 4.

We also discuss how to detect the presence of ecological fallacy when individual data are not available and concentrate on two approaches in particular: the fitting of models where transition probabilities are allowed to depend on covariates and bootstrap methods, see for instance \cite{Efron}.

The paper is organized as follows. In section 2 we define the setting and summarize the basic results concerning ecological fallacy and diagnostic tools to detect its presence. In section 3, after reconsidering the work of \cite{Liu07}, we investigate the possibility of correcting bias by modelling the effect of covariates; we also describe data generating mechanisms where modelling the effect of covariates is unlikely to make any improvement. In section 4 we describe the model of voting behaviour implicit in the modified version of the \cite{BrownPayne} method of ecological inference and analyse two sets of artificial electoral data affected by ecological fallacy.  In section 5 we present an application to the election for Mayor in the city of Turin in 2016 by combining a bootstrap simulation and an attempt to correct for ecological bias by fitting models with covariates.
\section{The ecological fallacy revisited}
Suppose that we are interested in voting transitions within a relatively small area like a town made of $K$ polling stations.  Let $Y$ denote the choice of a voter among the $C$ options available in a given election and $X$ the corresponding choice among the $R$ options that were available in a previous election. Let $p_{uij}$ denote the proportion of voters who chose $Y=j$ among those who voted $X=i$ in the previous election in polling station $u$. Let also $\pi_{ij}$ denote the probability that an eligible voter chosen at random among those who voted $X=i$ will chose $Y=j$; these quantities are the main object of ecological inference. They will be called {\it voter transitions} or {\it transition probabilities} though, when $i=j$, they measure the faithfulness of voters of party $i$; note that they sum to 1 by row.

Let also $y_{uj}$ and $x_{ui}$ denote, respectively, the aggregate proportions of voters for party $j$ in the new election and for party $i$ in the previous election in polling station $u$. By simple algebra, as for instance in \cite{Gnaldi18}, it can be shown that the usual ecological regression may be written as
$$
y_{uj} = \sum_1^R x_{ui} \pi_{ij} + \epsilon_{uj}, \text{ where }
\epsilon_{uj}=\sum_1^R x_{ui}(p_{uij}-\pi_{ij});
$$
a closer investigation of the nature of the error term $\epsilon_{uj}$ indicates that the condition for unbiased estimates by any methods based on linear or non linear regression on the marginal proportions $x_{ui}$ is that these are uncorrelated with the $p_{uij}$, the proportion of voter transitions within each polling stations. The effects of these correlations on bias depends on their signs and strengths; as explained by \cite{Wakefield}, section 3.3, in the $2\times 2$ case the largest biases are to be expected when correlations are all in the same direction. Predicting size and direction of bias in general is very complicated, some insights are provided in the example presented in section 4.

Basically, there will be ecological bias whenever the variations across polling stations of any of the proportions $p_{uij}$ is correlated with any of the marginal proportions $x_{ui}$. This is the {\it x-bar} rule of \cite{Firebaugh} where x-bar stands for averages which, in the context of voter transitions, are just the marginal proportions. Association between voting transitions and marginal proportions may be intrinsic, like when, for instance, the proportion of faithful voters of a given party is greater in those polling stations where the party was stronger in the previous election. Association may also be induced when certain entries of the table of voter transitions are correlated with an external variable which, in turn, is correlated with the marginal proportions; an example of this phenomenon is outlined in section 4.2.
\subsection{Robinson's data}
A rather complex set of multilevel models were fitted by \cite{Subra09} to a slightly extended version of the data used by \cite{Robinson}; though these data are not about voter transitions, they may be useful for clarifying the nature of the ecological fallacy. To test the condition for ecological fallacy in this context, note that the proportion of residents belonging to the different ethnic groups (Whites born in US, Whites not born in US and Blacks) are the analog of our $x_{ui}$, the proportion of votes obtained by each party in the previous election, while the proportion of illiterate within each ethnic group, are the analog of our $p_{uij}$; because $C=2$ and $p_{ui2}$ = $1-p_{ui1}$, we restrict attention to $p_{ui1}$. The issue is whether the variations by State of $p_{ui1}$, the proportions of illiterate, are correlated with those of the $x_{ui}$.

\begin{figure}
\centering
\includegraphics[width=13cm,height=6cm]{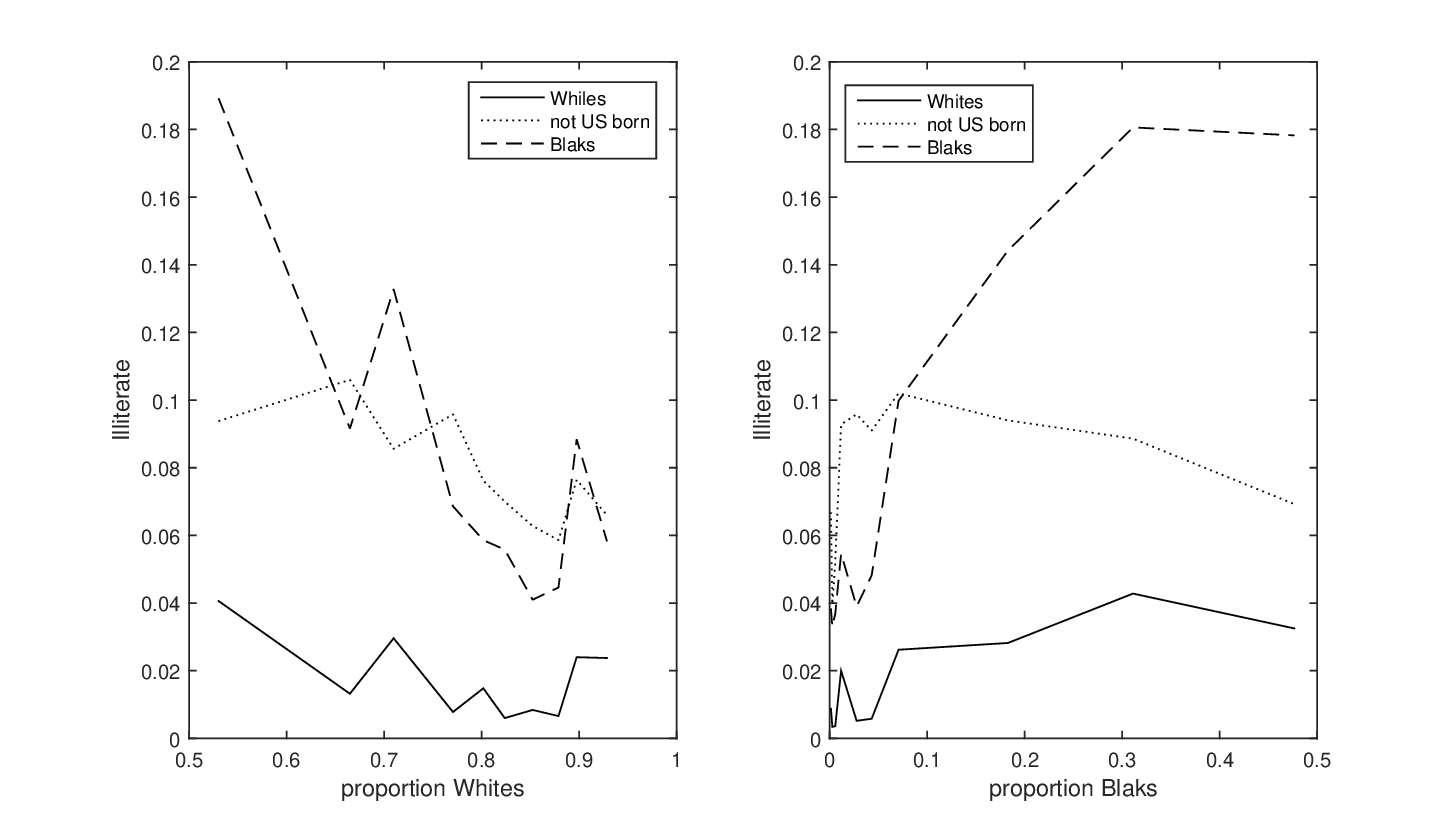} 
\caption{\label{Fig1} {\it Quantile regression of the proportion of illiterate by race against the proportion of Whites US born (left panel) and the proportion of Blaks (right panel).}}
\end{figure}

An informal assessment  may be obtained by examining the quantile regression plots of Figure \ref{Fig1}; these were obtained by classifying States in 10 groups according to the deciles of the explanatory variables (for instance the proportion of Black residents). Then, within each group of States, we plot the averages proportion of illiterates by race against the average proportion of Blacks (or Whites).

It emerges clearly that the proportion of illiterate decreases even among Blacks when the proportion of Whites increases; the relation is reversed, though with a different shape, when the proportion of Blacks increases. Robinson's data were also reanalysed by \citep[Section IV]{Firebaugh} who argued that region (South / rest of the Country) was the very determinant of illiteracy. In this example the proportion of Blacks is positively correlated with the proportion of illiterates within Blacks and Whites US born..
\subsection{Tools for detecting the ecological fallacy}
In the unlikely event that the full individual data were available, one could simply compute the $p_{uij}$ proportions and plot these as functions of the marginal proportions $x_{ui}$; for instance, for a give $i$ we might chose a set of $j$s of interest. A quantile regression as in Figure \ref{Fig1} may be used for a visual inspection of the overall direction of correlations.

When only the condensed individual data are available as in \cite{Plescia2018}, for each cell in the table of voter transitions, we can compute the errors in the estimates from ecological inference relative to the true values. However these errors, as in any statistical context, will be the sum of bias and random variation, unless we have substantial reasons to assume that ecological bias is negligible. in the given context

When only aggregated data are available, we cannot even know which estimates are closer to the truth. A set of indirect diagnostic tools for checking whether conditions for unbiased estimates are satisfied were discussed by \cite{Loewen1989} but criticized by \cite{Freedman1991}. In this paper we propose two diagnostic tools more directly related to the nature of ecological fallacy: (i) fitting models with covariates, (ii) comparison of standard errors estimated from theory with the corresponding bootstraps estimates. Point (i) will be the subject of next section; point (ii) is briefly described below and an application will be presented in section 5.

To detect ecological bias by bootstrap, a statistical method for obtaining non parametric estimates of standard errors \citep{Efron}, the chosen method of  ecological inference has to be applied several times to different subsets of polling stations selected at random with replacement. The idea is that, if transition probabilities are not constant across polling stations, the estimates obtained from different samples will exhibit an amount of variability larger than expected from theory. The method can be computationally demanding because the chosen method of ecological inference has to be applied, say, 300 times. In the end, 300 different estimates of each entry of the table of voter transitions will be available so that uncertainty in each estimate can be assessed without the need to make any assumption. An illustration will be provided in section 5.
\section{Ecological fallacy and covariates}
Several methods of ecological inference allow to model the effect of covariates on transition probabilities, for instance \cite{BrownPayne} and \cite{RoJiKiTa}; this possibility is also mentioned as a possible extension by \cite{GreinerQuinn}. However, as far as we know, nobody has noted that, only covariates which are correlated with the $x_{ui}$, the proportion of voters for the different options available in the previous election, are relevant for removing bias.

The assumption that the logits of transition probabilities can be a linear function of covariates measured at the level of polling station is stated by \cite{RoJiKiTa} in their equation (2); an identical assumption is formulated by \cite{BrownPayne} in their equation (2.3). Both formulations can be rewritten as
\begin{equation}
\log\frac{\pi_{uij}}{\pi_{uiC}}=\alpha_{ij}+\bd\beta_{ij}\tr \bd z_{uij},
\label{loreg}
\end{equation}
where $\alpha_{ij}$ is an intercept parameter, $\pi_{uij}$ is the transition probability from $i$ to $j$ in local unit $u$, $\bd z_{uij}$ is a vector of covariates affecting cell $(ij)$ and $\bd\beta_{ij}$ is a vector of regression coefficients acting on the logit scale.

Covariates measured at the level of polling stations which might affect voting decisions are, for instance, the proportion of people aged over 65, the proportion of unemployed or the proportion of immigrants. The idea behind the method is that, if voter transitions are not constant and their variation across polling stations are determined by certain covariates, which are correlated with the marginal proportions as in equation \ref{loreg}, then the induced ecological bias will be removed.

Individual data from the 2002 New Orleans mayoral election were used by  \cite{Liu07} to show that, by modelling the dependence of transition probabilities on appropriate covariates, one can reduce substantially the ecological bias. From a Statistical point of view this is not surprising because, generally, a correctly specified statistical model leads to consistent estimates. An example based on computer generated electoral data that confirms this expectation will be provided below.

A detailed discussion of social mechanisms by which individual behaviours may be affected by average features of a local unit (polling station) are discussed in depth by \citep[pp 564--568]{Firebaugh}. A specific examination concerning the way that the environment voters live in can affect their decisions is contained in
\cite{Johnston06},  in particular the chapter: {\it "Talking together, voting together"}.
\subsection{Individual level covariates}
According to \cite{Firebaugh}, in addition to covariates which, by social mechanisms, may affect groups of voters in a given polling station, covariates may also act as {\it micro properties}, when voters decide mainly on the basis of their personal condition (like unemployment, party affiliation or other), rather than the value of the same variable averaged at the level of their polling station.

A numerical example with artificial electoral data where voting decisions depend on individual covariates will be presented in section 4 and is based on the following assumptions. Suppose that the probability that an individual is unemployed varies across polling stations and that employed and unemployed had different probabilities of choosing among the voting options available in the previous election. Assume also that the probabilities of choosing among the options available in the new election depend both on the party voted previously and on the personal condition (employed or not).

In this context it is likely that the entries in the table of voter transitions will be correlated with both the marginal proportions in the previous election and with the proportion of unemployed. As we shall see, modelling the effect of covariates in this context may fail to correct ecological bias.
Intuitively, there are two reasons why this may happen: (i) as explained by \cite{Gnaldi18}, when decisions are affected by the personal condition (employed or not) transition probabilities are linear (not logistic) functions of the proportion of unemployed and (ii) to apply a proper ecological inference in this context we should know the proportion of unemployed within voters of different parties in the previous election, not just the average proportion of unemployed.
\section{Ecological fallacy in artificial electoral data}
In this section we describe a stochastic data generating mechanism for producing artificial electoral data having specific features; the analysis of these data will allow an empirical investigation into the ecological fallacy and  on models with covariates. The idea is to generate electoral data in a way that mimics voters' decisions in a context with a very large number of polling stations; this has two important advantages: (i) when the number of polling stations is very large, the random variation in the ecological estimates will be very small, so that bias, if present, is  easily detected, (ii) for the same reason, the experiments we present can be easily replicated  with the software that will be made available.
The data generating mechanism is derived from the modification of the \cite{BrownPayne} model proposed by \cite{FoGnBr}: in our opinion, the original paper and its modification, describe in detail a quite realistic stochastic data generating mechanism which can be implemented on a computer.

The artificial data will be analysed with three methods of ecological inference: the Goodman linear regression, the frequentist version of the Multinomial-Dirichelet method by \cite{RoJiKiTa} and the modified \cite{BrownPayne} method itself. In principle,  the last method might be favoured by the fact that the data are generated according to the model underlying the method; however, due to the size of the data, random variation will be negligible relative to bias and here the objective is not to compare the three methods, but to show that, under certain conditions, ecological fallacy is going to affect any method, approximately in the same way.
\subsection{The revised Brown and Payne model of voting behavior}
The approach to ecological inference proposed by \cite{BrownPayne} is based on a stochastic model of voting behaviour which may be seen as a simplified, but also realistic, approximation of reality. The model assumes the presence of two random components: (i) at the individual level, the decision about which party to vote in the new election is assumed to be similar to the drawing of balls from an urn with a given composition, determined by the party voted in the previous election and the polling station where the individual lives; (ii) at the level of polling station, it is assumed that transition probabilities for voters of the same party may vary at random across polling stations due to local specificities which are too complex to be known and are treated as random effects.

The revised version of this model due to \cite{FoGnBr} assumes that voters of the same party and living in the same polling station are not homogeneous and that they tend to split into small clusters of people connected by personal relationships. To be specific, the model assumes that the size of clusters is determined at random and only people within the same cluster share the same set of transition probabilities. In turn, these cluster specific transition probabilities vary at random as in a Dirichlet distribution, a model of random variation of sets of probabilities which is also part of the \cite{RoJiKiTa} model. In the final step, the parameters of the Dirichlet distributions, one for each voting option available in the first  election, are allowed to depend on covariates measured at the level of polling stations.

An interesting feature of this model is that it can be used to design a computer algorithm that generates at random electoral data with preassigned characteristics. These data can be analysed both as full, individual level data and as aggregated data to gain a better understanding of the ecological fallacy by computer experiments. The objective of these experiments is not to compare the performance of different methods of ecological inference, but to show that, when the basic conditions are violated in a given way, biased estimates are to be expected no matter which method of estimation is used. As an illustration, we concentrate on three method of ecological inference: \cite{Goodman} linear regression model, the frequentist version of \cite{RoJiKiTa} Multinomial-Dirichlet and the revised Brown and Payne model.

Our approach will consist in generating artificial data sets with 20,000 polling stations each. The data generating mechanism depends on the following set of parameters which have to be set in advance:
\begin{enumerate}
\item a set of transition probabilities and a model that specifies which transition probability depend on which covariate; recall that, because covariates which are uncorrelated with the marginal proportions $x_{ui}$ are irrelevant, we may restrict the range of relevant models;
\item the amount of random variation in the Dirichlet distribution which determines the amount of heterogeneity about the behaviour of voters belonging to different clusters;
\item the average size of clusters, a parameters that has to be specified but has little effect on the functioning of the model.
\end{enumerate}
\subsection{Examples}
Below we consider two examples, in the first we assume that covariates affect the logits of voter transitions as expected by both the \cite{RoJiKiTa} and the \cite{BrownPayne} models, in the second, instead, covariates are assumed to act as micro properties. In the first setting it emerges clearly that, when covariates are ignored, all three methods of inference provide estimates heavily biased; however the bias disappear when covariates are inserted properly into the model. On the contrary, in the second example modelling the effect of covariates does not remove bias, as explained above.

In the first example we suppose that the same three voting options are available in both elections and that voters split among these in equal proportions in the first election. We also assume that the logits of the transition probabilities within each row are linear functions of the corresponding marginal proportion in such a way that faithfulness to the option voted in the first election is greater in the polling stations where the same party was stronger in the first election. As we shall see, this leads regression based methods to overestimate the amount of faithful voters.
\begin{table}[ht]
\caption{\label{tab:1} True and ecological estimates of voter transitions in example 1: T=true, K=King, B=Brown-Payne and G=Goodman; C stands for model with covariates; U, V, Z denote the three voting options available in each election}
\centering
\fbox{

\begin{tabular}{lrrrcrrrcrrr}
Method & \multicolumn{3}{c}{T} & \hspace{1mm} & \multicolumn{3}{c}{KC} & \hspace{1mm} & \multicolumn{3}{c}{BC}  \\
 \hline
 & U & V & Z & & U & V & Z & & U & V & Z \\
U &  0.714 & 0.143 & 0.143 & & 0.715 & 0.145 & 0.140 & & 0.712 & 0.145 & 0.142\\
V &  0.142 & 0.716 & 0.142 & & 0.141 & 0.714 & 0.144 & & 0.142 & 0.713 & 0.144\\
Z &  0.142 & 0.142 & 0.715 & & 0.142 & 0.141 & 0.716 & & 0.144 & 0.142 & 0.714\\ \hline
Method & \multicolumn{3}{c}{G} & \hspace{1mm} & \multicolumn{3}{c}{K} & \hspace{1mm} & \multicolumn{3}{c}{B}  \\
 \hline
 & U & V & Z & & U & V & Z & & U & V & Z \\
U & 1.000 & 0.000 & 0.000 & & 1.000 & 0.000 & 0.000 & & 0.936 & 0.033 & 0.031\\
V & 0.000 & 1.000 & 0.000 & & 0.000 & 1.000 & 0.000 & & 0.031 & 0.937 & 0.031\\
Z & 0.000 & 0.001 & 0.999 & & 0.000 & 0.000 & 1.000 & & 0.033 & 0.031 & 0.936\\

\end{tabular}}
\end{table}
On the other hand, errors in the ecological estimates with covariates do not exceed 0.0024, a value compatible with random variation; estimates without covariates are all heavily biased in the same direction: the values along the main diagonal (proportions of faithful voters) are much larger than the truth while transitions towards other parties are underestimated.

In the second example, for simplicity, we assume that there are only two voting options in both elections, leading to  a $2\times 2$ table and assume that there is a binary variable, for instance unemployment, which affects the probability that a voter chooses one of the two available options in the first election. This induces correlation between the party chosen in the first election and the proportion of unemployed. In addition, in the second election, we assume that voters' decisions depend both on the party voted in the previous election and on whether they are employed or not.
\begin{table}[ht]
\caption{\label{tab:2} True and ecological estimates of voter transitions in example 2: T=true, K=King, B=Brown-Payne and G=Goodman; C stands for model with covariates}
\centering
\fbox{

\begin{tabular}{lrrcrrcrr}
Method & \multicolumn{2}{c}{T} & \hspace{1mm} & \multicolumn{2}{c}{KC} & \hspace{1mm} & \multicolumn{2}{c}{BC}   \\
 \hline
 & U & V & & U & V & & U & V  \\
U & 0.2583 & 0.7417 & & 0.0461 & 0.9539 & & 0.1399 & 0.8601\\
V & 0.1462 & 0.8538 & & 0.3588 & 0.6412 & & 0.2648 & 0.7352\\
\hline
Method & \multicolumn{2}{c}{G} & \hspace{1mm} & \multicolumn{2}{c}{K} & \hspace{1mm} & \multicolumn{2}{c}{B}   \\
 \hline
 & U & V & & U & V & & U & V  \\
U & 0.4042 & 0.5958 & & 0.4104 & 0.5896 & & 0.4043 & 0.5957\\
V & 0.0001 & 0.9999 & & 0.0000 & 1.0000 & & 0.0000 & 1.0000  \\

\end{tabular}}
\end{table}
Here estimates without covariates with the three methods are all heavily biased in a similar way. As covariates we used both the proportion of voters for party V in the first election and the proportion of unemployed and it turns out that the estimates are again far away from truth, though in a different direction; in addition both methods behave similarly.
\section{An application to real data}
This study concerns the second ballot in the 2016 Municipal elections in Turin, one of the largest urban area in Northern Italy,  Though in the Italian system, since 1993, Municipal elections are based on a two-round system, during the two most recent elections in Turin, held in 2006 and 2011, the candidate of the centre-left coalition was elected in the first round. Considering the traditional strength of the centre-left coalition in Turin and that, due to personal rivalries, the parties on the centre-right were running with three different candidates, most political analysts had predicted that candidate Piero Fassino, incumbent mayor and one of the Democratic Party (PD) leader at the national level,  would probably win in the first ballot.

Since 2013 the Italian political scenario is considered to be a tripolar party system, with  the Movimento 5 Stelle (M5S) having gained a strength comparable to that of the more traditional centre-left and centre-right coalitions, a feature which has, gradually, extended to several local political systems, see for example \cite{Emanuele} and \cite{Regalia}. Though the M5S was active in Turin since the 2011  Municipal election, only after the 2016 Municipal election M5S became the third-largest political group in the local scenario; the M5S candidate was Chiara Appendino, former party chief councillor. On the whole there were six main candidates running at the first ballot; the remaining ones were grouped into two residual categories:
\begin{enumerate}
\item Airaudo, (LEFT), heading a left-wing post communist coalition;
\item Fassino, (PD), heading a centre-left candidate;
\item Appendino, (M5S), candidate of the Five-Stars Movement;
\item Napoli, (FI), running for Forza Italia, Berlusconi's party;
\item Morano, (LN), Lega Nord’s candidate;
\item Rosso (UDC), another centre-right party;
\item (ODX), other candidates on the right;
\item (OSX), other candidates on the left;
\item (NOV), in both ballots this category denotes abstainer, blank votes and other unclassified votes
\end{enumerate}

The 2016 first-round election took place on June 5th 2016 and, surprisingly, the centre-left leader did not win as expected and Fassino (PD) and Appendino (M5S) entered into a runoff election which was held two weeks later. Though the expected winner was again the centre-left coalition leader, he was clearly defeated, a result that represents a true turning point in the Turin local political history. The objective of the estimation and analysis of voter transitions from the first to the second ballot tries to throw some new light on what has happened. One conjecture that we aim to verify is that a substantial amount of voters of centre-right candidates excluded from the ballot voted M5S rather than abstaining, which would be an instance of strategic voting \citep{Herrmann}, when voters make the effort to go to the ballots to prevent the victory of the candidate they disliked most.

Turin has 909 valid polling stations, after removing hospitals, prisons and the like, with an average of 786 voters each.
To account for possible heterogeneity in voting behaviour, we used a geographic information system software (GIS) to determine the position of the 909 polling stations and link to the most relevant social-economic variables from Census data. By matching polling stations with census data, the following five covariates, measured at the level of polling stations and expressed as ratios with respect to the number of eligible voters, were made available: aged 65 and over, registered immigrants, unemployed, low qualified workers and people with low education.
\begin{figure}
\centering
\makebox{\includegraphics[width=14cm,height=7.5cm]{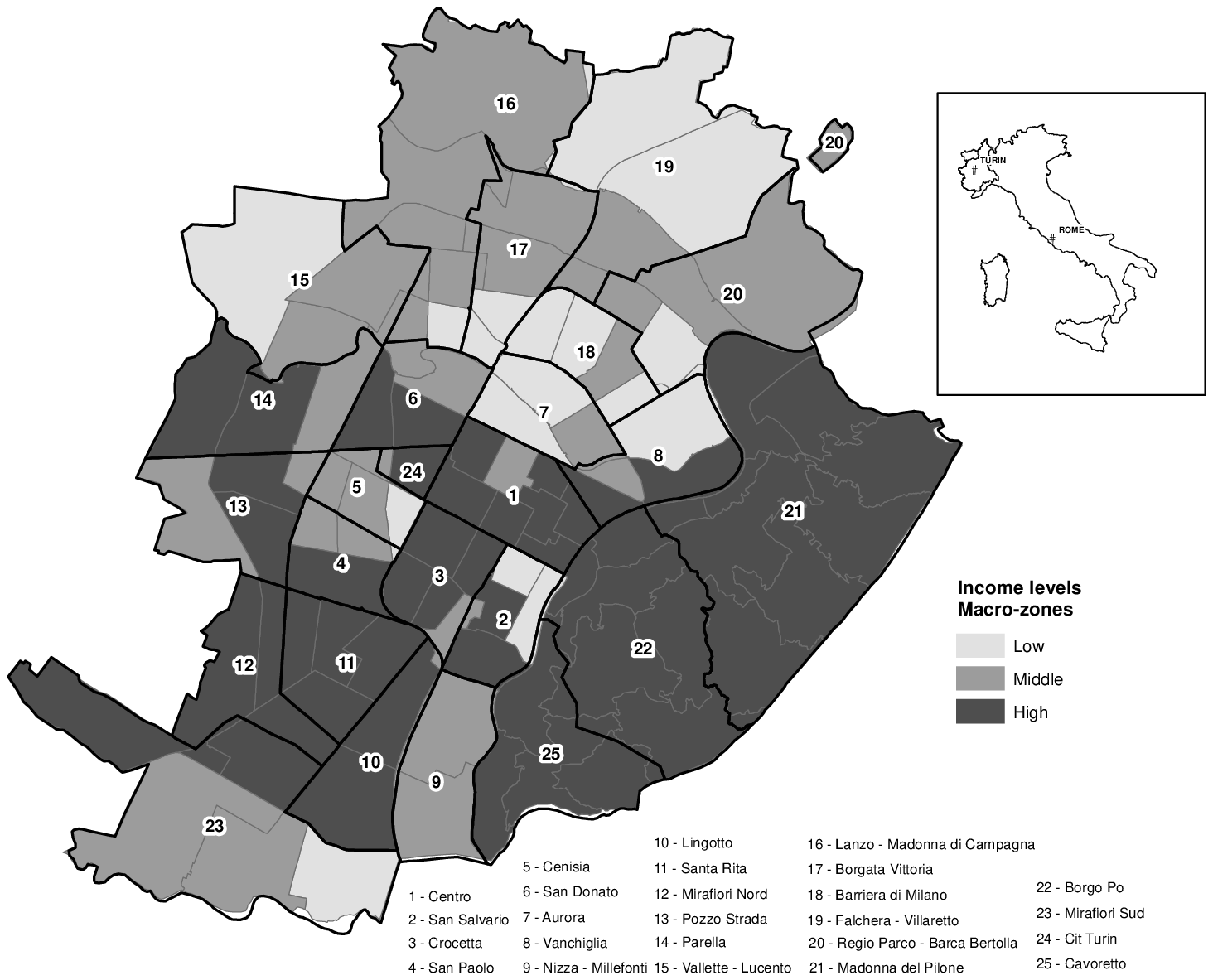}}
\caption{\label{Fig2} {\it Map of zones in Turin borough, darker area correspond to higher average income}.}
\end{figure}
We started by fitting a modified Brown and Payne model without covariates and then tried several alternative models which allow the most relevant cells of the table of voter transitions to depend on the available covariates. An informal model selection procedure was used by comparing the estimates of the various regression coefficients on the logit scale with the corresponding standard errors and removing non-significant covariates. Evidence that the final model with covariates is a substantial improvement is provided both by the substantial increase in the log-likelihood from about -11800 to -5680 and the fact that all estimated regression coefficients are significant at the 5\% level. Estimates of transitions provided by the model without covariates and by the final model are displayed in Table \ref{tab:3}.

\begin{figure}
\centering
\includegraphics[width=13cm,height=7.5cm]{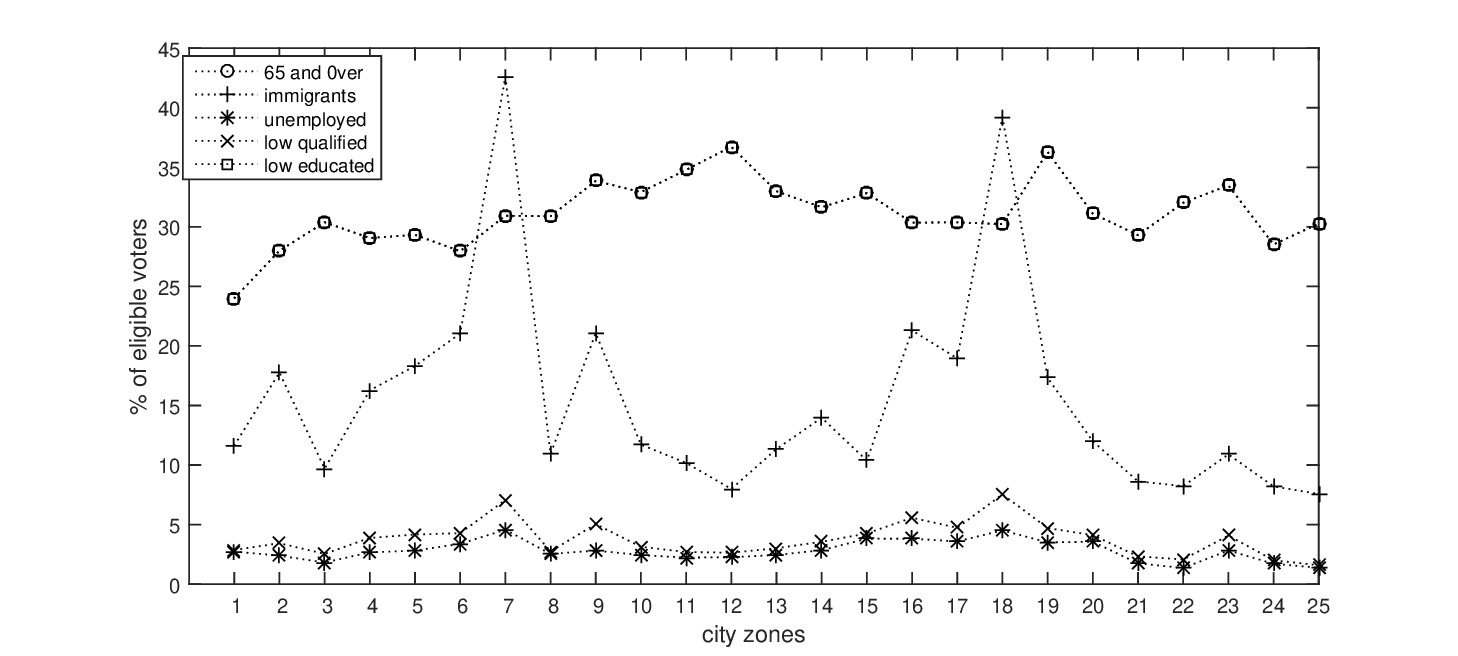} 
\caption{\label{Fig3} {\it Average covariate values within the 25 macro areas composing the Municipality of Turin.}}
\end{figure}
\begin{table}[ht]
\caption{\label{tab:3} Estimates of voter transitionsbetween the first and the second ballot in the city of Turin with and without covariates}
\centering
\fbox{
\begin{tabular}{lrrrcrrr}
 & \multicolumn{7}{c}{Second Ballot} \\
First Ballot & \multicolumn{3}{c}{without covariates} & \hspace{1mm} & \multicolumn{3}{c}{with covariates}  \\
 & M5S & PD & NOV & & M5S & PD & NOV \\
 \hline
LEFT & 0.000 & 0.894 & 0.106 & & 0.093 & 0.841	& 0.066 \\
M5S  & 1.000 & 0.000 & 0.000 & & 1.000 & 0.000 & 0.000 \\
UDC  & 1.000 & 0.000 & 0.000 & & 0.947 & 0.000 & 0.053 \\
FI   & 0.828 & 0.096 & 0.076 & & 0.797 & 0.203 & 0.000 \\
PD   & 0.033 & 0.944 & 0.023 & & 0.036 & 0.902 & 0.061 \\
LN   & 0.808 & 0.096 & 0.096 & & 0.754 & 0.153 & 0.092 \\
ODX  & 1.000 & 0.000 & 0.000 & & 0.479 & 0.000 & 0.521 \\
OSX  & 0.496 & 0.000 & 0.504 & & 0.600 & 0.189 & 0.211 \\
NOV  & 0.005 & 0.000 & 0.995 & & 0.025 & 0.006 & 0.969 \\

\end{tabular}}
\end{table}
Estimates of the regression coefficients and the corresponding ratios $z$ between estimates and standard errors for the model with covariates are displayed in Table \ref{tab:4}. The absolute value of the $z$ ratio gives a measure of significance in the sense that, the greater the value, the more unlikely it is that the corresponding association is a random artifact; usually values greater that 1.96 are taken as significant. To interpret the results one should consider the sign of the regression coefficients together with the cell of the table of voter transitions and the nature of the corresponding covariates.
\begin{table}[ht]
\caption{\label{tab:4} Estimated regression coefficient for the model with covariates: An = preoportion of voters ages 65 and over; Im = proportion of immigrants; Bs = proportion of residents with low schooling, Bq = proportion of low qualified workers }
\centering
\fbox{
\begin{tabular}{lrrrrlrrrr}
Cov & row & col & est. & $z$ &  Cov & row & col & est. & $z$  \\
 \hline
An& 1&	2&	0.59&	3.20&	Bq&	4&	1&	-0.85&	-2.98\\
Im&	1&	2&	0.44&	3.21&	An&	5&	2&	-0.01&	-2.08\\
Bs&	1&	2&	-0.96&	-3.19&	Bs&	6&	1&	0.16&	4.63\\
An&	4&	1&	-0.14&	-4.61&	Di&	6&	1&	-0.50&	-3.92\\
Im&	4&	1&	-0.17&	-3.09&	Im&	7&	1&	-0.21&	-4.13\\
Bs&	4&	1&	0.92&	5.27&	Bq&	7&	1&	0.42&	2.37\\
\hline
\end{tabular}}
\end{table}

Additional evidence that the estimates of voter transitions with covariates provide a closer approximation to the truth is provided by the comparison of the estimates of the standard errors of voter transitions according to theory (assuming no ecological bias) and by bootstrap, for both the models without and with covariates. The results are displayed in Table \ref{tab:5} where the first row is the sum by columns of the differences between the bootstraps standard error (Bse) and the theoretical standard error (Tse) for the cells where Bse is greater than Tse; the second row contains again the sum by columns of the differences (Tse - Bee) this time for the cells where Tse is greater.

Note that, for the model without covariates, the values in the first row are substantially larger than those in the second row which are close to 0; this indicates that Tse underestimate substantially the true standard errors probably because transition probabilities vary across polling stations due to ecological fallacy. On the other hand, the corresponding results for the model with covariates indicate that Tse is approximately equal to Bse and that their differences are small and go in both directions, as if they were caused by random variations.
\begin{table}[ht]
\caption{\label{tab:5} Differences between standard errors expected from theory (Tse) and estimated by bootstrap (Bse), columns totals.}
\centering
\fbox{
\begin{tabular}{lrrrcrrr}
  & \multicolumn{3}{c}{Without covariates} & \hspace{1mm} & \multicolumn{3}{c}{With covariates}  \\
 & M5S & PD & No vote & & M5S & PD & No vote \\
 \hline
$Bse-Tse,\: Bse>Tse$  & 0.329 &0.073 &0.338& & 0.043& 0.049	&0.040\\
$Tse-Bse,\: Tse>Bse$  & 0.008 &0.001 &0.000& & 0.034& 0.023	&0.071\\
\hline
\end{tabular}}
\end{table}

In other words, Table \ref{tab:5} tries to provide a summary of the relevant differences between standard errors expected from theory and resulting in reality as estimated by bootstrap. The fact that the values of the first row are substantially larger relative to the second indicates that the estimates provided by the corresponding model are likely to suffer  from ecological bias.
\section{Concluding remarks}
One of the main conclusions of this paper is that, whenever the proportion of voters transitions are not constant across polling stations and their variations are correlated with the strength of certain parties in the previous election, estimates from aggregated data will be biased, no matter which method of ecological inference is used. Though methods of ecological inference differ in the way they model the randomness intrinsic to voters' decisions, when the number of available polling stations is relatively large, this component is likely to be almost negligible relative to potential ecological biases. Only methods that allow to model the effect of covariates measured at the level of polling stations have a chance to correct or reduce the bias.

By modelling the effect of covariates on transition probability, we are implicitly assuming that covariates act as {\it macro properties}, to use the terminology of \cite{Firebaugh}, meaning that certain average features of polling stations affect the decisions of those who voted in the same way in the previous election. A detailed study of situations where this may happen are discussed by \cite{Johnston06}. However, there may be situations where voters' decisions depend on their personal condition (e.g. young, unemployed) rather than on the corresponding proportions within the polling station: that is we are faced with {\it micro properties}. When this is the case, it is no longer true that modelling covariates is going to reduce ecological bias. We describe a specific data generating mechanism and give a numerical example where modelling the effect of covariates fails to correct the bias.

We also argue that, unless data at the individual level are available, it will ba almost impossible to find objective evidence of ecological fallacy. However, when covariates measured at the level of polling stations are available, indirect evidence of bias may be achieved by fitting models with covariates. To be specific, when fitting a model which allows transition probabilities to depend on a given covariates, we may check how much better does this model fit relative to the one without covariates and test whether the estimated regression coefficients are all highly significant. When both these conditions are satisfied, it is very unlikely that the detected dependence of transitions from covariates is a numerical artifact.

An additional technique which we consider for detecting ecological bias without having access to individual data is based on comparisons between the standard errors in the estimated transitions derived form theory with the corresponding standard errors estimated by bootstrap. Essentially, the bootstrap consists in computing estimates of voter transitions from several random samples of polling stations. When the conditions for no ecological fallacy are violated, by selecting at random different subsets of polling stations, we are likely to obtain estimates which differ more than what would be expected by the randomness intrinsic in voters' choices in a homogeneous context.

Finally, we would like to comment on the linear regression method of ecological inference, also known as the Goodman method, which, because of its computational simplicity, is widely used in certain contexts. Because this method cannot account for the effect of covariates, it should be applied only within an area which is sufficiently small to believe that transition probabilities do not differ in a systematic way across polling stations. For instance, the results presented in the previous section suggest that, though Goodman's method should not be applied to the city of Turin as a whole, its application should probably be valid within suitably chosen sub-areas of the Municipality.  In doing so, we need to balance two conflicting requirement: (i) the number of polling stations should not be too small, otherwise estimates will be affected by large random fluctuations; (ii) if the area is too large, it is likely that estimates are affected by a certain amount of ecological bias. An advantage of the Goodman method which, as far as we know, has not yet been exploited, is that a bootstrap estimate of standard errors of voting transitions can be computed very easily and compared with those expected in an ideal situation.

\bibliographystyle{spbasic}
\bibliography{QuaQua}
\end{document}